\documentclass[twocolumn,preprintnumbers,nofootinbib,aps]{revtex4}
\usepackage{epsfig,amssymb,amsmath}

\newcommand{\bi}{\bibitem}
\newcommand{\be}{\begin{eqnarray}}
\newcommand{\ee}{\end{eqnarray}}
\newcommand{\rar}{\rightarrow}

\begin{document}

\title{Natural extension of the Generalised Uncertainty Principle}

\author{C.~Bambi$^{\rm 1}$}
\author{F.R.~Urban$^{\rm 2,3,4}$}
\affiliation{$^{\rm 1}$Department of Physics and Astronomy, Wayne State University, Detroit, MI 48201, USA\\
$^{\rm 2}$Istituto Nazionale di Fisica Nucleare, Sezione di Ferrara, I-44100 Ferrara, Italy\\
$^{\rm 3}$Dipartimento di Fisica, Universit\`a degli Studi di Ferrara, I-44100 Ferrara, Italy\\
$^{\rm 4}$Department of Physics and Astronomy, University of Sussex, Brighton BN1 9QJ, UK}

\date{\today}

\preprint{WSU-HEP-0703}

\begin{abstract}
We discuss a gedanken experiment for the simultaneous measurement 
of position and momentum of a particle in de Sitter spacetime. 
We propose an %further 
extension of the so called 
Generalised Uncertainty Principle (GUP) which {\it implies} the 
existence of a minimum observable momentum. 
The new GUP is directly connected to the non--zero cosmological 
constant, which becomes a necessary ingredient for a more complete 
picture of the quantum spacetime. 
\end{abstract}

\maketitle

{\sc Introduction ---} In quantum mechanics, two operators which do not commute 
cannot be measured simultaneously with arbitrary 
accuracy. In the case of position $x$ and momentum along 
the same direction $p$, we obtain the well known formula
\be\label{up}
\Delta x \, \Delta p \gtrsim 1 \, ,
\ee
where $\Delta x$ and $\Delta p$ denote position and 
momentum uncertainty respectively (we use $\hbar = c = 1$ 
units throughout the paper). 
When one takes gravitational interactions into account, 
eq.~(\ref{up}) must be revised, since the classical 
concept of spacetime breaks down as we approach distances 
close to the Planck length $L_{Pl} \sim 10^{-33}$ cm. 
Even if at present we do not have a fully reliable and trustworthy theory of 
quantum gravity, from rather general and more or less model 
independent considerations~\cite{gups, maggiore, a-s} one 
can conclude that the Uncertainty Principle~(\ref{up}) 
should be replaced by the so called Generalised 
Uncertainty Principle (GUP)
\be\label{gup}
\Delta x \, \Delta p \gtrsim 1 
+ \alpha \, L_{Pl}^2 \, (\Delta p)^2\, ,
\ee
where $\alpha$ is a positive dimensionless parameter which 
is expected to be of order one. It is easy to see that 
eq.~(\ref{gup}) implies the existence of a minimum observable 
length, $(\Delta x)_{min} \approx 2 \, \alpha^{1/2} \, L_{Pl}$.
Among other considerations, this could be interpreted as an 
indication of the necessity of replacing point--like particles 
with extended objects in any consistent theory of quantum 
gravity~\cite{maggiore}, and as a signal of the breakdown of the concept 
of continuum spacetime at very small scales~\cite{a-s}. Of course, 
by turning off gravitational interactions eq.~(\ref{gup}) reduces 
to eq.~(\ref{up}), since $L_{Pl} \rar 0$.

However, one notes that eq.~(\ref{gup}) is neither very 
``appealing'' from an aesthetic point of view, nor ``democratic'',
because the position $x$ and the momentum $p$ seem to play 
different roles: the square of $\Delta p$ appears on the right 
hand side of eq.~(\ref{gup}), but the square of $\Delta x$ 
does not. Therefore, on such grounds, one is tempted to 
try to consider possible extensions 
of eq.~(\ref{gup}). The most natural one is
\be\label{pre-sgup}
\Delta x \, \Delta p \gtrsim 1 
+ \alpha \, L_{Pl}^2 \, (\Delta p)^2 \, 
+ \beta \, \frac{(\Delta x)^2}{L_X^2} \, ,
\ee
where $\beta$ is a new dimensionless order one coefficient 
and $L_X$ a new unknown fundamental length. This extended GUP is invariant under the following transformations
\be\label{symm}
\Delta x &\rar& \left(\frac{\alpha}{\beta}\right)^{1/2} \,
( L_X L_{Pl} ) \Delta p \, ,\nonumber\\
\Delta p &\rar& \left(\frac{\beta}{\alpha}\right)^{1/2} \,
( L_X L_{Pl} )^{-1} \Delta x \, ,
\ee
which explicitely underline the symmetry properties of eq.~(\ref{pre-sgup}).

Such a proposal, with $L_X = L_\Lambda$, where $L_\Lambda = (3/\Lambda)^{1/2}$ is the de Sitter horizon, had been already put forward in ref.~\cite{cavaglia}, in order to obtain heuristically the temperature of (Anti--) de Sitter black holes along the same line of reasoning which permits the derivation of the Hawking temperature for Schwarzschild black holes from the standard Uncertainty Principle~(\ref{up}). There are, however, two important differences.

First, in ref.~\cite{cavaglia} a {\it positive} cosmological constant gives rise to a {\it negative} $\beta$, thereby preventing the interpretation of the extended GUP as an indication for the existence of a minimum observable momentum.

Second, the argument of ref.~\cite{cavaglia} has been criticised in~\cite{scardigli}, where it is shown that (Anti--) de Sitter black hole temperature can also be deduced using only eq.~(\ref{up}).

In this letter, by performing a simple gedanken experiment where we aim at measuring position and momentum of a particle in de Sitter spacetime, we reobtain eq.~(\ref{pre-sgup}) with $L_X = L_\Lambda$, but now with a {\it positive} $\beta$: $\beta > 0$. This result differs from the previous one and, importantly, implies the existence of a theoretical bound on the minimum momentum we can measure. In particular, here we will use only quantum mechanics and general relativity, while in ref.~\cite{cavaglia} eq.~(\ref{pre-sgup}) arises from high energy physics and the quantum structure of spacetime\footnote{Something similar happens for instance in the derivation of the GUP. In~\cite{gups} the framework is that of string theory, whereas in ref.~\cite{a-s} the author relies on only quantum mechanics and general relativity arguments.}.

A final remark for this introduction: a relation between minimum observable momentum and cosmological constant was also conjectured from group theory arguments in the context of deformed special relativity theories in ref.~\cite{livine}. Indeed, just from the definition of expectation value and unceratinty of an observable, we have
\be
(\Delta A)^2 (\Delta B)^2 \ge 
\frac{1}{4} \left| \langle [A,B]\rangle \right|^2
\ee
where $A$ and $B$ are two general observables (in our case they would be $x$ and $p$). So, there is always a strong connection between unceratinty relations and commutation relations.
%Indeed, in general different commutation relations (between position and momentum operators in our case), and therefore different underlying algebras, lead to different uncertainty relations~\cite{livine}. 
These hypothesis, however, were not followed by supportive operational derivation. Moreover, the heuristic argument we present here, although less rigorous, is more general and cautious, because it does not require a well defined framework and, as it is known, the choice of the commutation relation leading to a particular uncertainty relation is not unique. The relation between different choices are not yet completely clear~\cite{livine}.

\begin{figure}[t]
\par
\begin{center}
\includegraphics[width=8cm,angle=0]{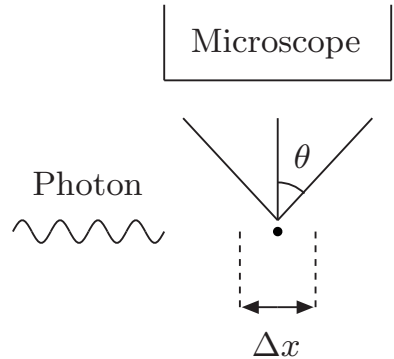}
\end{center}
\par
\vspace{-7mm} 
\caption{Setup of the experiment. Since we do not know the 
exact position of the electron, when $\Lambda \neq 0$ we 
cannot know the exact cosmological redshift of the photon. 
This leads to an extra contribution to the momentum 
uncertainty $\Delta p$.}
\label{fig}
\end{figure}

{\sc The standard GUP ---} Let us now recall very briefly the standard experiment of 
the Heisenberg microscope, in which we want to measure position 
and momentum of a particle, say an electron. The setup of 
the experiment is sketched in fig.~\ref{fig}. Using a photon 
with wavelength $\lambda$, we are able to resolve 
at best length scales of order $\lambda$ itself, 
so the projected electron position 
uncertainty is $\Delta x \gtrsim \lambda/\sin\theta$, where 
$\theta$ is the angle defined in fig.~\ref{fig}. At the same time, the 
interaction between the electron and the photon changes the original 
momentum of the former resulting in $\Delta p \gtrsim \sin\theta/\lambda$. 
Thus, we arrive at eq.~(\ref{up}).

With the same experimental 
setup, we can also obtain directly eq.~(\ref{gup}), for 
example if we replace the electron with an extremal black 
hole~\cite{maggiore}. The latter absorbs the photon with 
wavelength $\lambda$ and then decays back to the extremal 
state, emitting a single 
photon with the same wavelength (extremal black holes 
are believed to be stable and thermal description breaks down 
for near extremal states). Before absorption and after emission,
the black hole mass is $M$ and its radius is $R$, whereas in 
the time interval between the two events the mass is 
$M' = M + 2 \, \pi \, \lambda^{-1}$ and the radius is
\be
R' \gtrsim R + 2 \, G_N \, \lambda^{-1} \, .
\ee
Since the measurement is unable to discriminate between 
the two radii, and 
$G_N = L_{Pl}^2$, we can conclude that gravity introduces 
an extra contribution to position uncertainty, 
i.e. $\Delta x_{new} \gtrsim L_{Pl}^2 \, \lambda^{-1}$. 
The lower bound on $\Delta x$ in term of 
$\Delta p \sim \sin\theta/\lambda$ is eq.~(\ref{gup}).

This result has been found using a rather complicated, though explicit, setup, but it is important to notice how we could have expected it very simply on dimensional grounds. Indeed, in addition to the standard uncertainty one can reasonably argue that gravitational interaction between the photon and the particle should (weakly but inevitably) contribute. This term is independent on the nature of the probe, as follows from the Equivalence Principle. Hence, the leading order correction is expected to be proportional to $G_N$, which straightforwardly, for dimensional reasons, fixes the overall structure of~(\ref{gup}).

One even further simple and intuitive derivation is provided for example in~\cite{scardi2}. In quantum mechanics, the accuracy on the position of a particle is limited by the uncertainty on its momentum by $\Delta x \gtrsim 1/\Delta
p$. On the other hand, in general relativity we cannot localise a certain amount of energy in a region smaller than the one defined by its gravitational radius, i.e. $\Delta x \gtrsim G_N \Delta p$. If we combine the two result we conclude that
\be
\Delta x \gtrsim \max (1/\Delta p \, , \, G_N \Delta p) \nonumber\, .
\ee
The GUP then follows, upon multiplication by $\Delta p$.

{\sc The extended GUP ---} Now we assume to live in a de Sitter universe with cosmological 
constant $\Lambda$. Here the photon momentum is affected by the 
cosmological redshift, its magnitude depending on the 
path length of the photon before its detection: in particular, 
we should expect that an uncertainty on such a path length 
implies an uncertainty on the cosmological redshift and hence 
on $p$. The setup of the experiment is the same as before, the only difference is that 
the spacetime has a positive cosmological constant $\Lambda$.

Let us start with a very rough estimate, based on Newtonian 
mechanics, which reminds one of the approaches suggested 
in ref.~\cite{a-s} for the derivation of the GUP~(\ref{gup}) 
and would like to show that the result comes from rather 
general considerations on the effects of a non--zero 
cosmological constant. Indeed, like in the GUP, the key-point 
is the validity of the Equivalence Principle, which implies 
that the phenomenon is independent of the experimental setup,
that is of the nature of the particle and of the probe. 
Since the framework is 
non--relativistic, we replace the photon with a non--relativistic 
particle of mass $m$. It is easy to see that the Newtonian 
acceleration induced by a particle of mass $M$ on a second 
one in a spacetime with cosmological constant $\Lambda$, is (for 
more details on the derivation of this limit, see e.g.\ ref.~\cite{newton})
\be\label{e-newton}
\ddot{\bf r} = \left( -\frac{G_NM}{r^2} 
+ \frac{\Lambda}{3} \, r \right) 
\, \frac{{\bf r}}{r} \, ,
\ee
where ${\bf r}$ is the position vector of the second particle 
with respect to the one of mass $M$. The formula is obtained 
considering the Newtonian limit in static and spherically 
symmetric coordinates. Of course, eq.~(\ref{e-newton}) 
reduces to standard Newtonian gravity for $\Lambda \rar 0$.
From eq.~(\ref{e-newton}) we see that in our gedanken experiment 
there is an inevitable effective extra acceleration, due to 
a non--zero cosmological constant. This new contribution will affect the interaction 
between the electron, whose position and momentum we want 
to measure, and the particle we use as a probe, in particular by making the momentum of the outgoing particle uncertain. Relying once again on dimensional analysis we can infer the structure of the leading correction to be proportional to $\Lambda \sim L_\Lambda^{-2}$, and therefore to $(\Delta x)^2$.

The outlined Newtonian estimate may appear quite 
unreliable, or significant only in a particular limit; 
let us now consider a more sophisticated and complete picture. 
We will work in Friedman--Robertson--Walker metric with zero spatial curvature 
\be\label{frw}
ds^2 = dt^2 - a(t)^2\left( dr^2 + r^2 d\Omega^2 \right) \, ,
\ee
where $a \sim \exp(Ht)$ is the cosmological scale factor. 
Assuming that we know exactly the electron position 
along the direction orthogonal to $x$, we send a photon 
towards the electron along the $x$ direction, as shown in 
fig.~\ref{fig}. Since the position $x$ of the point where 
the photon hits the electron and is then scattered, is 
known with a non--zero uncertainty $\Delta x$, the uncertainty 
in the path length of the photon before its detection by 
the microscope is at best at the same level, that is, 
$\Delta x$. This implies that the momentum of a photon 
is affected by the redshift logarithmic uncertainty $\Delta q / q$
\be\label{deltap}
\frac{\Delta q}{q} = \frac{\Delta\lambda}{\lambda} = \frac{\Delta a}{a} = H \, \Delta t \approx H \, \Delta x = \frac{\Delta x}{L_\Lambda}\, ,
\ee
where\footnote{We can replace $\Delta t$ by $\Delta x$ only in the small $H \Delta t$ limit, which is the limit of interest given today's small $H$.} $H = (\Lambda/3)^{1/2}$ is the Hubble parameter in the de Sitter universe. Notice that the redshift acts on the momentum of the photon, not on the wavelength, whereas the astrophysical redshift is defined operatively through the latter. Here we recast the equation by means of the de Broglie wavelength in order to give a more intuitive description of the phenomenon, but the use of $\lambda$ is not necessary in obtaining the final result.

Moreover, note that even though $H$ is usually defined through the total energy density, i.e. the energy density of matter plus the one associated to $\Lambda$, here we take into account only the cosmological constant contribution, the reason being that matter energy density is spatially constant only on large scales, but most universe is basically empty and, if $\Lambda = 0$, one can always perform the experiment in a static spacetime, for example in a Schwarzschild vacuole of a Friedmann universe, where there is not expansion at all (for more details, see e.g.\ section 27.3 of ref.~\cite{stephani}). On the other hand, a true cosmological constant is spread everywhere and is present at all scales, so the experiment subjects always to the expansion of the universe. In other words, the expansion produced by a matter density can be seen as an accidental source of systematic error, depending on where the experiment is performed; the expansion caused by $\Lambda$ instead is an intrinsic property of the spacetime and is inevitable.

Now, the point here is that since we do not know the exact interaction point we equivalently do not have arbitrary precision on the redshift of the photon, which then translates into a momentum uncertainty, eq.~(\ref{deltap}). We also stress here that the zeroth order relation (\ref{up}) still holds, since
\be
\Delta x \simeq \lambda_T \exp^{H(t_M-t_T)}/\sin\theta \, ,
\ee
but also
\be
\Delta q \simeq \sin\theta \exp^{-H(t_M-t_T)} / \lambda_T \, ,
\ee
where $\lambda_T$ is the wavelength as measured at the interaction point $T$ (the subscripts $M$ and $T$ referring to quantities as measured at the microscope and at the interaction point respectively). This means that we are allowed to add to the old uncertainty principle the newly found uncertainty.

If we use the fact that $\Delta x \gtrsim \lambda$ as follows from~(\ref{up}), and that the errors are then to be added by quadrature, the low energy Uncertainty Principle reads
\be\label{result}
\Delta p \gtrsim \Delta q \gtrsim 
\sqrt{\frac{1}{(\Delta x)^2} + \frac{1}{L_\Lambda^2}} \simeq \frac{1}{\Delta x} + \frac{\Delta x}{2 \, L_\Lambda^2} + \dots \, ,
\ee
thereby implying the Extended Uncertainty Principle (EUP)
\be\label{eup}
\Delta x \, \Delta p \gtrsim 1 
+ \beta \, \frac{(\Delta x)^2}{L_\Lambda^2} \, .
\ee

Let us gather what we have so far: by combining~(\ref{eup}) 
with the high energy source of uncertainty one arrives at 
the formula
\be\label{new}
\Delta x \, \Delta p \gtrsim 1 
+ \alpha \, L_{Pl}^2 \, (\Delta p)^2
+ \beta \, \frac{(\Delta x)^2}{L_\Lambda^2} \, ,
\ee
where $\beta$ is a positive dimensionless coefficient which 
cannot be deduced from our simple considerations. The reson for this is that we have used several inequalities and trivial simplifications, although it is natural to expect this coefficient to be of order one - we comment on this point in the conclusions.
Notice further that, as it must be, when $\Lambda \rar 0$ and the de Sitter universe 
becomes the Minkowski spacetime, eq.~(\ref{new}) reduces 
to eq.~(\ref{gup}).

In writing this result we have actually taken advantage of some external arguments/hints. First of all, the overall structure of the eq.~(\ref{new}) is inferred by dimensional (model independent) arguments, the leading order corrections being proportional to $G_N$ and $\Lambda$. This is a very general, intuitive, and reasonable expectation.

Secondly, there is a substantial difference in the derivations of the GUP~(\ref{gup}) and the EUP~(\ref{eup}). Indeed, the black hole argument leading to~(\ref{gup}) makes use of two separate and distinct gedanken experiments for the standard and high energy contributions. The way uncertainties are combined, that is, linearly, is dictated by dimensional arguments and/or group theory ones, by picking up a particular extension of the Poincar\'e algebra. The expression for the EUP instead results from considering one single ideal experiment, and expanding up to leading order, thereby requiring the standard statistical error treatment (quadrature). This can be traced in our statement that $\Delta x \gtrsim \lambda$, which is true only by means of the standard uncertainty relation, which we found to be valid in de Sitter spacetime.

Moreover, if we do not want to rely on the standard uncertainty priciple for the derivation of the EUP, following~\cite{scardi2}, we should write
\be\label{alternate}
\Delta p \gtrsim \max ( 1 / \Delta x \, , \, p \Delta x / L_\Lambda ) \, ,
\ee
and therefore
\be\label{p-minE}
\Delta x \, \Delta p \gtrsim 
1 + \frac{(\Delta x)^2}{L_\Lambda} \, p \gtrsim 
1 + \frac{(\Delta x)^2}{L_\Lambda} \, p_{min} \, ,
\ee
where now $p_{min}$ is the minimum observable momentum which we
now try to estimate. Since $p_{min} \gtrsim \Delta p$,
we can write
\be
\Delta x \, p_{min} \gtrsim  
1 + \frac{(\Delta x)^2}{L_\Lambda} \, p_{min}
\ee
and then
\be\label{p-minE2}
p_{min} \gtrsim \frac{L_\Lambda}{\Delta x \, L_\Lambda 
- (\Delta x)^2} \, .
\ee
Of course $p_{min} \gtrsim 0$, so $\Delta x \lesssim L_\Lambda$.
Moreover, the minimum of eq.~(\ref{p-minE2}) is for 
$\Delta x \sim L_\Lambda/2$, which implies $p_{min} \gtrsim 4/L_\Lambda$,
and we get the following expression for the new term of momentum uncertainty
\be\label{result+}
\Delta p \gtrsim \frac{\Delta x}{L_\Lambda^2} \, ,
\ee
which, having been obtained independently, can be linearly added to the standard Heisenberg uncertainty.

By analysing eq.~(\ref{new}) one is led to conclude about the minimum observable momentum. However there is a subtlety here, which is traceable in our derivation. Eq.~(\ref{deltap}) knows nothing about the possible existence of $(\Delta p)_{min}$, and it holds for $\Delta q / q < 1$, i.e.\ $\Delta x / L_\Lambda < 1$. The same assumption is required in the Newtonian framework as well, because in order to write eq.~(\ref{e-newton}) we need $r^2/L_\Lambda^2 \ll 1$. On the other hand, in the end we assume that the new uncertainty principle has a wider range of validity and we get the prediction that there exists a minimum observable momentum for $\Delta x \sim L_\Lambda$. Let us note that a similar trick affects the standard GUP in~(\ref{gup}), where the minimum observable position is gotten for $\Delta p \sim M_{Pl}$, but we cannot be sure that the formula still holds and does not break down earlier (for example, in some theoretical frameworks single particles cannot exceed the Planck energy). Indeed, one may expect that both the additional terms to the standard Heisenberg Uncertainty Principle~(\ref{up}) are only the first order correction of a more general formula which can be possibly deduced only from the true quantum theory of gravity. On the other hand, this is the drawback of all the heuristic considerations which try to investigate something beyond their realm of validity. Still, the qualitative picture should be trusted.

It is important to note that we obtain the same result even in 
static coordinates, where there is no true cosmological redshift. 
For example, if we take the de Sitter metric in static and 
spherically symmetric coordinates, where the line element is 
\be
ds^2 = \left(1 - \frac{r^2}{L_\Lambda^2}\right) dt^2
- \frac{dr^2}{1 - \frac{r^2}{L_\Lambda^2}}
- r^2 d\Omega^2 \, ,
\ee
the photon redshift and blueshift depend only on the point of the 
spacetime where the photon is detected, regardless of where 
the collision between the photon and the electron takes place. 
Despite that, here we have an horizon $L_\Lambda$ and the photon 
wavelength cannot exceed the de Sitter horizon, i.e. the 
effective radius of the spacetime. Hypothesising that there are no 
dramatic changes of the ordinary de Broglie relation up to 
wavelength of order $L_\Lambda$, we find $q_{min} \sim 1/L_\Lambda$, 
which implies and justifies the introduction of the last term of eq.~(\ref{new}) 
into the usual GUP.

{\sc Conclusions ---} We conclude with a few comments. First of all, from this point of view 
the cosmological constant $\Lambda$ appears as a fundamental 
ingredient of the spacetime, at the same level of the gravitational 
constant $G_N$, without which we cannot regard 
position $x$ and momentum $p$ as lying on common grounds. 
Here we would like to emphasise once more that in our approach $\beta > 0$, yielding for the extraction of a minimum observable momentum, and restoring the symmetry between position and momentum broken by the usual GUP. 
Such an idea is even reinforced by 
present astrophysical and cosmological observations which favour a tiny 
but non--zero $\Lambda$ responsible for the present accelerating 
expansion rate of the universe~\cite{data}. On the other hand, 
the cosmological constant remains an unexplained free parameter 
to be determined by experiments.

As for possible physical implications of eq.~(\ref{new}), 
one expects departures from the canonical commutation 
relations in quantum mechanics and in quantum field theory and, 
as by--product, consequences in thermodynamics and statistical 
mechanics. The new GUP may be seen as an indication 
of quantum gravity deviations from the classical spacetime 
at both very small and very large scales (present cosmological 
observations suggest the value $L_\Lambda \sim 10^{28}$ cm); 
this may also be consistent with studies on the stability 
of Minkowski spacetime~\cite{stability}. 
IR deviations from classical general relativity would 
represent alternative signatures in our search for evidence of 
the quantum structure of the spacetime and may be easier to 
observe than the expected UV phenomena near the Planck scale. In this connection, let us notice that the actual practical applications appear quite unlikely, which in turn implies that there are not stringent significative experimental limits on the coefficient $\beta$: the numbers we are dealing with are too small.
The new picture may also (probably: only) play an important role in the physics 
of the early universe, during the inflationary epoch~\cite{inflation}, 
where the effective cosmological constant had to be much larger 
than the one in the present universe, with $L_\Lambda$ probably 
at the level of $10^{-28} - 10^{-24}$ cm. However in this case a more accurate derivation of eq.~(\ref{new}) is demanded, as some of our expansions and simplifications do no longer apply. We plan to study some consequences of the Extended GUP in another work.

\begin{acknowledgments}
We would like to thank Alexander Dolgov, Alexey Petrov, David Bailin, Stephan Huber, Anke Knauf, and Stephanie Stuckey for helpful comments.
F.U. wishes to thank the Particle Theory Group at the University of Sussex for kind hospitality while this work was being written.
C.B. is supported in part by NSF under grant PHY-0547794 and by DOE under contract DE-FG02-96ER41005.
F.U. is supported by INFN under grant n.10793/05.
\end{acknowledgments}

\end{document}